\documentclass[aps,amsmath,amssymb,preprintnumbers,nofootinbib,a4paper,prd,twocolumn]{revtex4-1}
\pdfoutput=1
\usepackage{amsthm}
\usepackage{graphicx}
\usepackage{color}
\usepackage{bbm}

\usepackage{mathrsfs}
\usepackage{amssymb}
\usepackage{amsthm}
\usepackage{graphicx,amsmath}

\newcommand{\be}{\begin{eqnarray}}
\newcommand{\ee}{\end{eqnarray}}
\newcommand{\bea}{\begin{eqnarray}}
\newcommand{\eea}{\end{eqnarray}}

\usepackage{epsfig}

\definecolor{rossoCP3}{cmyk}{0,.88,.77,.40}

\begin{document}

\title{\color{black} Asymptotically Safe Dark Matter} 
\author{Francesco Sannino}
\email{sannino@cp3-origins.net} 
\author{Ian M. Shoemaker}
\email{shoemaker@cp3-origins.net} 

%\affiliation{
%\vspace{5mm} 
%{ \color{rossoCP3}  \rm CP}$^{\color{rossoCP3} \bf 3}${\color{rossoCP3}\rm-Origins} \& the 
% {\color{rossoCP3} \rm Danish IAS} 
%\mbox{ University of Southern Denmark, Campusvej 55, DK-5230 Odense M, Denmark}}
\affiliation{{\color{black} CP$^{3}$-Origins} \& Danish Institute for Advanced Study {\color{black}, Danish IAS}, University of Southern Denmark, Campusvej 55, DK-5230 Odense M, Denmark}
 \begin{abstract}
We introduce a new paradigm for dark matter (DM) interactions in which the interaction strength is asymptotically safe. In models of this type, the coupling strength is small at low energies but increases at higher energies, and asymptotically approaches a finite constant value. The resulting phenomenology of this ``asymptotically safe DM'' is quite distinct. One interesting effect of this is to partially offset the low-energy constraints from direct detection experiments without affecting thermal freeze-out processes which occur at higher energies.  High-energy collider and indirect annihilation searches are the primary ways to constrain or discover asymptotically safe dark matter.   \\[.1cm]
%{\footnotesize  \it Preprint: XXX}
 %\preprint{CP3-Origins-2014-047 DNRF90, DIAS-2014-47}
%\vskip -.1cm
{\footnotesize  \it Preprint: CP$^3$-Origins-2014-047 DNRF90, DIAS-2014-47} 
 \end{abstract}

\maketitle

%\section{Running matters} 
\textbf{\textit{Running Matters -}} Significant theoretical and experimental effort is underway in an effort to unveil the fundamental nature of the non-luminous component of matter. While very little is known about this dark matter (DM), evidence for its existence is overwhelming, coming from multiple strands of inquiry.  One of the few properties of DM that is very well known is its cosmological abundance: $\Omega_{CDM}h^{2} = 0.1199 \pm 0.0027$~\cite{Ade:2013zuv}.  One well-studied framework for understanding the relic abundance of Dark Matter (DM) is thermal freeze-out~\cite{zeldo}. Number-changing interactions in the early universe, $\overline{X}X \leftrightarrow (\overline{{\rm SM}}) {\rm SM}$, keep DM in thermal equilibrium with the Standard Model (SM) bath, until the rate of these annihilation processes drops below the rate of Hubble expansion. After this point the abundance of DM is essentially fixed, with a value scaling as $\Omega_{DM} \propto 1/\langle \sigma_{ann}v_{rel}\rangle$, where $\langle \sigma_{ann}v_{rel}\rangle \simeq 6 \times 10^{-26}~{\rm cm}^{3}~{\rm s}^{-1}$ gives the observed DM abundance.  One of the most stringent constraints on DM of this type come from direct detection (DD) experiments which search for the nuclear recoil deposited in a detector from a DM scattering event, $X N \rightarrow X N$. Yet, the absence of any unambiguous DD signal indicates that many of the simplest models for the thermal relic abundance are already ruled out. 

%Today, a wealth of experimental probes are placing strong constraints on this {well-motivated} scenario from the null observations of its annihilation~\cite{Ackermann:2013yva}, production at colliders~\cite{CMSmonojets}, and nuclear scattering at direct detection experiments~\cite{Akerib:2013tjd}.  

We must however be careful when applying these experimental constraints to given particle physics models and it is incumbent upon us to re-examine the assumptions of their relevance for the thermal relic paradigm.  %For example, the experimental constraints on DM interactions come from conditions spanning several orders of magnitudes in energy. 
For example, importantly the process relevant for the relic abundance is DM annihilation which probes energies $2m_{X}$ while in DD the energies probed are $\mathcal{O}({\rm MeV})$. Thus in models where the DM mass is significantly heavier than an ${\rm MeV}$ there is a natural separation of scales that are relevant for annihilation and DD. In this present paper, we aim to exploit this fact in models with significant running in order to ease the tension between DD and the thermal relic hypothesis. 
%while direct detection experiments are very sensitive to the strength of the coupling of DM with ordinary matter involving momenta exchange $\mathcal{O}({\rm MeV})$, the relevant energy for indirect detection experiments which are sensitive to DM annihilation is twice the DM mass which is typically in the 1 GeV to 100 TeV range (by unitarity~\cite{Griest:1989wd}). %Often these constraints become even stronger once the assumption is made that the DM relic density is either thermal in nature or due to an asymmetry. This is so since one can further relate the coupling to ordinary matter to its mass. 

We know quite generally that couplings run when quantum corrections have been taken into account.  %and their variation can be substantial over the energy range spanned by current experiments in search of DM. 
A time-honored example is Quantum Chromodynamics (QCD). Here the squared coupling decreases by a factor of five when passing from a $\mathcal{O}({\rm GeV})$ to $\mathcal{O}({\rm 100~GeV})$. %Not only does the theory experience a large variation in the value of the coupling but the entire spectrum undergoes dramatic changes from a quark-gluon-like picture to a hadronic one at low energies.  

%Imagine that the DM interaction strength are such that they are small at low energies and grow at high energies in such a way to (partially) offset the direct detection experiments constraints but retaining all the other high-energy properties. 

%%%%%%%%%%%%%%%%%%%%%%%%%%%%%%%
\begin{figure*}[t] %  figure placement: here, top, bottom, or page
\begin{center}
 \includegraphics[width=.31\textwidth]{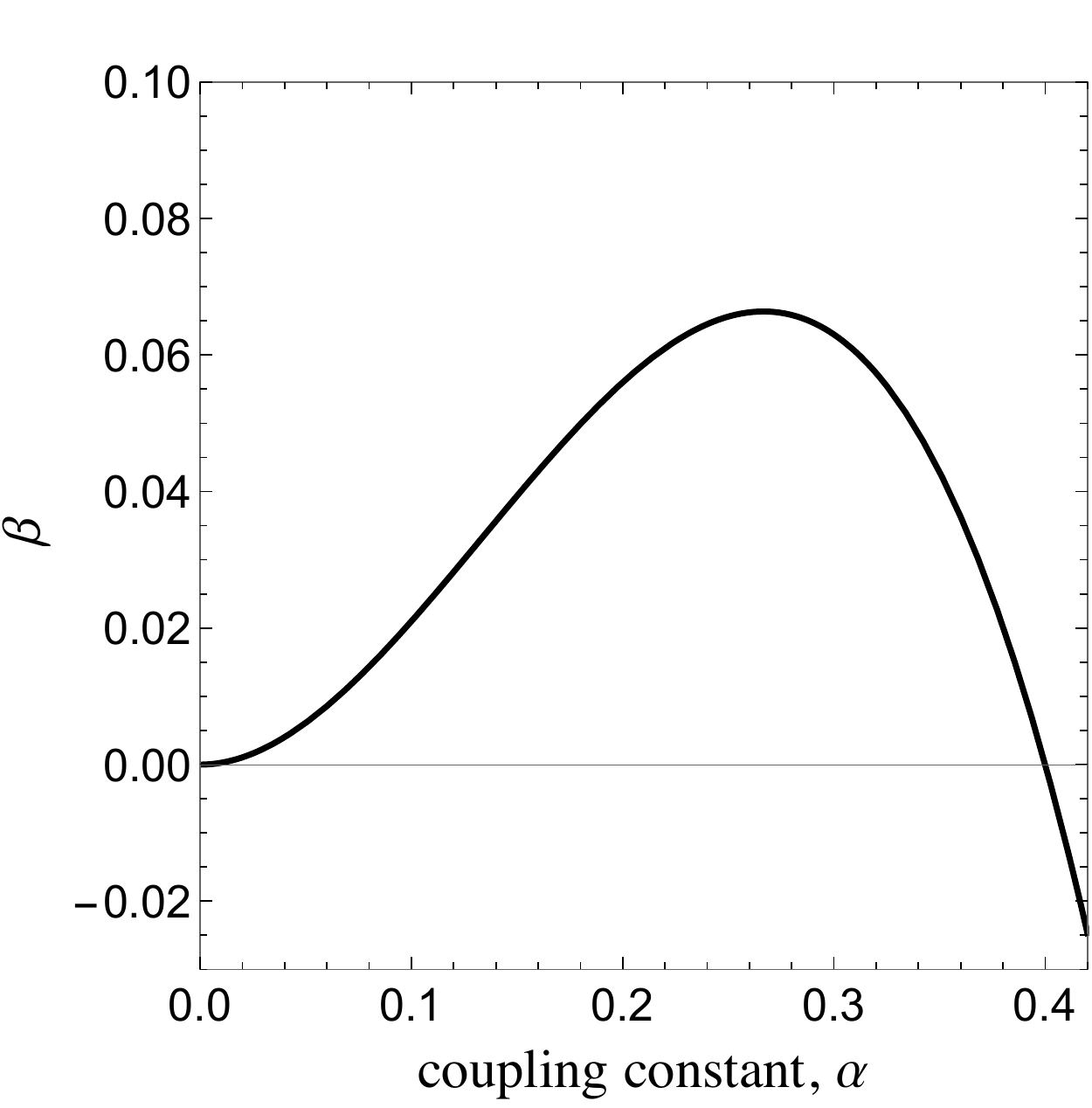} ~~~
  \includegraphics[width=.31\textwidth]{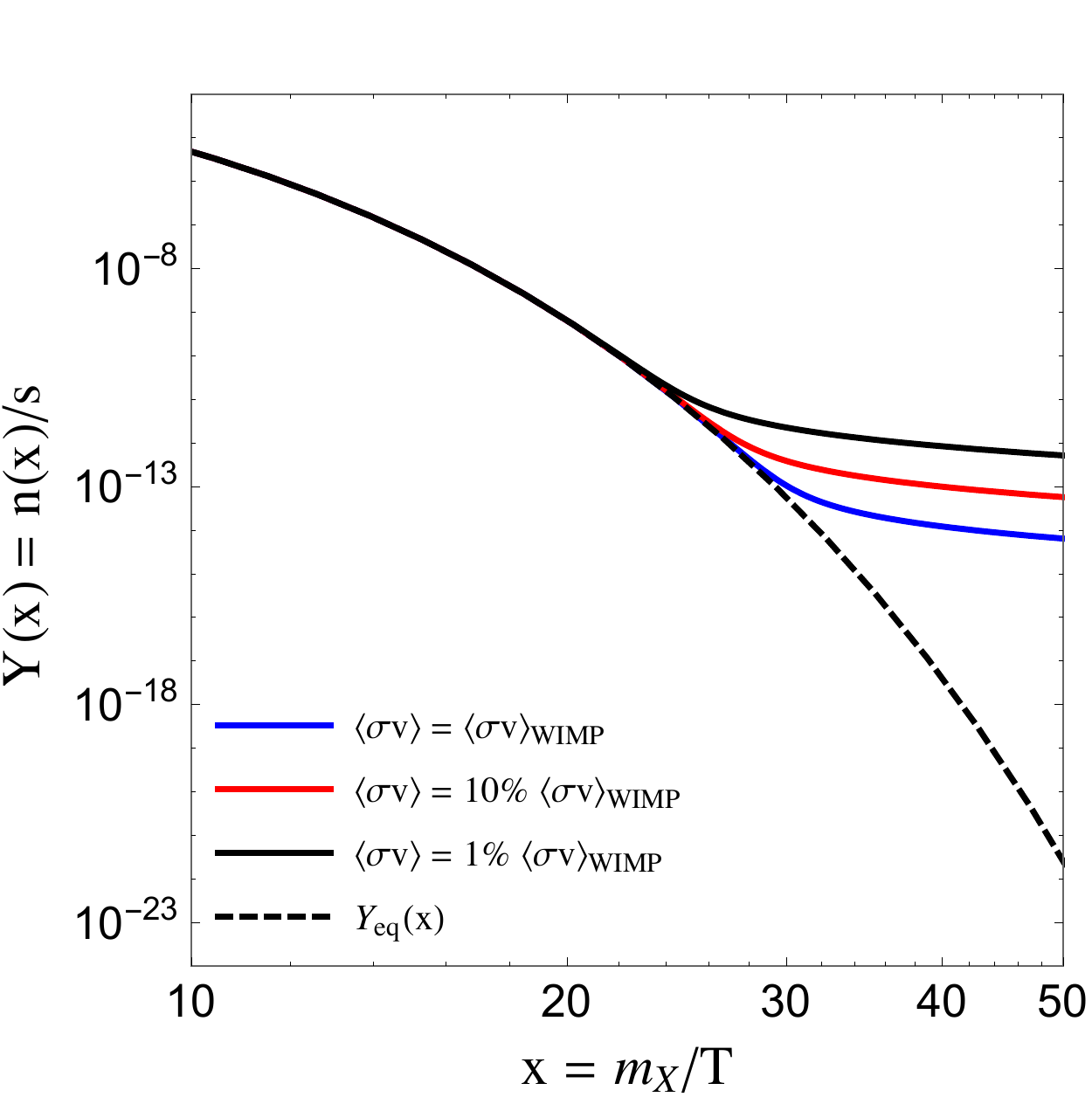} ~~~
  \includegraphics[width=.31\textwidth]{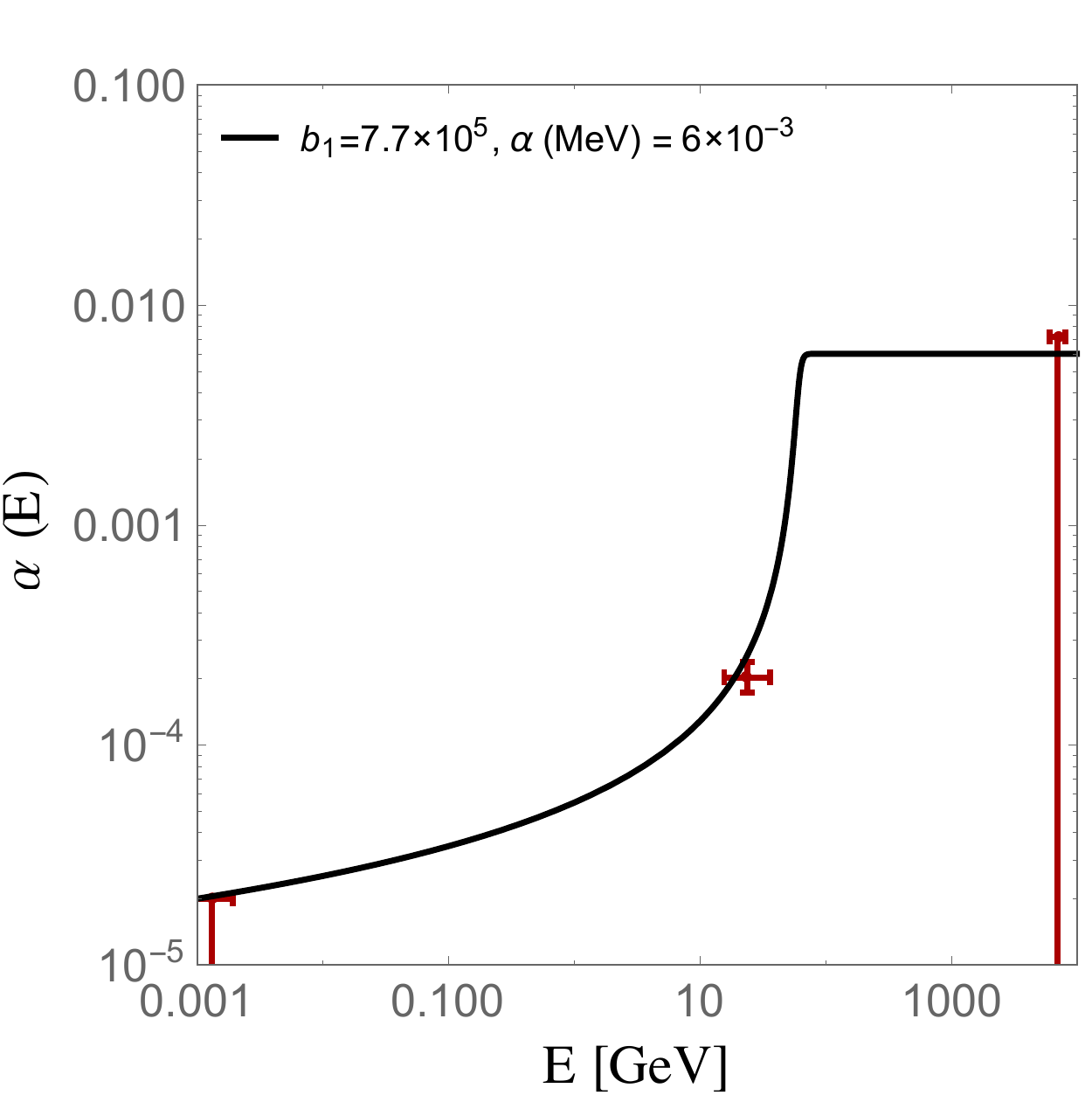} ~~~

\caption{{\it Left panel}: Asymptotically safe beta function given in Eq.~\eqref{ASFBeta} for $\alpha^{\ast} = 0.4$ and $b_1=7$. 
{\it Center panel}: Verification of standard freeze-out in asDM for annihilation cross sections near $\langle \sigma v \rangle_{{\rm WIMP}} = 3 \times 10^{-26}~{\rm cm}^{3}~{\rm s}^{-1}$.
{\it Right panel}: Energy dependence of the coupling  $\alpha(E)$. In this panel, we fix the mass and annihilation cross section to reproduce the best-fit point of the gamma-ray Galactic Center excess and take the mediator mass to be much less than the DM mass. In addition, we display the LUX and LHC monojet limits. }  % {\it Right panel}: Dependence of the intrinsic energy scale $\Lambda_{AF}$ on $b_1$ for $\alpha^{\ast} = 0.4$ and $\alpha(\mu_0)=0.04$ for $\mu_0 = 1$ MeV. Above $\Lambda_{AF}$ the coupling has already reached over $2/3$ of its fixed point value. }
\label{betafig}
\end{center}
\end{figure*}

%%%%%%%%%%%%%%%%%%%%%%%%%%%%%%%

%%%%%%%%%%%
%\begin{figure*}[ht!]
%	{  \includegraphics[width=0.4\textwidth]{betaAF_v2}}
%%\hfill
%	{ \includegraphics[width=0.4\textwidth]{running_v2}} 
%
%	{ \includegraphics[width=0.4\textwidth]{LambdaAF_v3}} 
%	
%		\caption{{\it Left}: Asymptotically safe beta function  given in Eq.~\eqref{ASFBeta} for $\alpha^{\ast} = 0.4$ and $b_1=7$. {\it Right}: Energy dependence of the coupling for $\alpha(\mu_0) = 0.04$ at the reference scale $\mu_0$.} \label{betafig}
%\end{figure*}
%%%%%%%%%%%

This is a convenient place to pause and reflect on the kind of coupling structure we will consider in the following. For simplicity, we imagine a dark sector connected to the SM via the exchange of a messenger particle. We shall take the messenger particle to be a vector, $V_{\mu}$, but expect similar results for scalars. The interactions between DM and the visible sector will be parametrized by a Lagrangian
\begin{eqnarray}
\mathscr{L}_{V} &=&  i\bar{X} \gamma^{\mu}\left(\partial_{\mu} - i g_X V_{\mu}  \right) X \nonumber \\ &&+ \bar{q} \gamma^{\mu}\left(\partial_{\mu} - i g_q V_{\mu}  \right) q + m^2_V V_{\mu}V^{\mu}\ . 
\label{eq1}
\end{eqnarray}
Here the interactions of the dark sector with the messenger is given by the coupling $g_{X}$ while the coupling of the messenger with the standard model (SM) quarks by $g_{q}$.  They both run with energy.  We stress that this simplified model of DM-SM interactions is adopted merely to illustrate the phenomenological implications of safe DM and more general constructions are possible. Simplified models of this type have become quite common in phenomenological studies since they contain the key parametric dependencies encountered in a large class of Beyond Standard Model constructions (see e.g.~\cite{Alves:2011wf}). 

With this setup, the cross sections for interaction with ordinary matter and DM annihilation are 
\begin{equation}
\sigma \propto  \frac{\alpha_{q} \alpha_{X} }{m_V^4} \mu^{2}  \ , \quad \langle \sigma_{\rm ann}v \rangle \propto  \frac{\alpha_{q} \alpha_{X}}{m_V^4}m_{X}^{2} \ ,
\end{equation}
with $m_V$ the mass of the messenger and $\alpha_{i} = g_{i}^2/4\pi$ where $i=q,X$. For simplicity we have assumed the interaction with ordinary matter to be identified with the SM quarks, and assumed the hierarchy $m_V > m_X$. The crucial point is that, because of the running of the couplings (due to the DM independent dynamics, and the dynamics of the messenger sector with ordinary matter), these cross sections can depend sensitively on the energy at which they are employed.  Crucially, the energies probed by direct detection and DM annihilation are typically different  by many orders of magnitude.

%For example at keV energies the cross section $\sigma$ can be identified with the proton cross section $\sigma_p$ which is essential for direct detection experiments. We can specialize $\sigma$ also for computing the DM annihilation cross section. This energy can be several order of magnitudes higher than the one relevant for direct detection experiments.  Depending on the underlying model one or both couplings can substantially change within the phenomenologically relevant energy range.

In contrast with QCD, we here take the underlying DM theory to be {\it asymptotically safe} rather than asymptotically free~\cite{Litim:2014uca}, and additionally assume that the couplings are always in the perturbative regime. This means that the couplings grow with energy towards the ultraviolet and become constant above a certain energy threshold that we call {$\mu_{S}$}. A simple beta function parametrization, for a generic coupling $\alpha$, that in four dimensions captures the essence of a non-gravitational asymptotically safe theory is: 
\begin{equation}
\mu\frac {d\alpha}{d{\mu}} = \beta = b_0 \alpha^2 - b_1 \alpha^3  = b_1 \alpha^2 (\alpha^{\ast} - \alpha)\ ,
\label{ASFBeta} 
\end{equation}
with positive $b$ coefficients and $\mu$ the energy (renormalization) scale. This beta function possesses two independent zeros. A non-interacting one for $\alpha = 0$ and an interacting one for
\begin{equation} 
\alpha^{\ast} = \frac{b_0}{b_1} \ .
\end{equation}
The coefficient $b_1$ partially controls how fast, in renormalization group time $t=\ln (\mu/\mu_0)$, the fixed point $\alpha^{\ast}$ is reached. Here $\mu_0$ is a reference energy corresponding to a given (theoretical or experimental) value of the coupling $\alpha(\mu_0) = \alpha_0$. 
For illustration we show the beta function in the left panel of Fig.~\ref{betafig} for the choice $\alpha^{\ast} = 0.4$ and $b_1=7$. 

Next, let us illustrate an interesting phenomenological implication of the asDM. The solution of the differential equation yielding the specific running for the coupling is exhibited in the right panel of Fig.~\ref{betafig} with the further assumption $\alpha_0 = 0.04$, yielding $\alpha^{\ast}/\alpha_0 = 10$.

It is phenomenologically relevant to investigate the dependence of the intrinsic scale $\mu_{S}$ above which the coupling has almost reached the ultraviolet fixed point. A simple definition we adopt is the energy scale above which the running coupling has reached $2/3$ of its fixed point value $\alpha^{\ast}$. %We show in the right panel of Fig.~\ref{betafig} the dependence of $\Lambda_{AF}$ on $b_1$ having fixed $\alpha(\mu_0) = 0.04$, $\alpha^{\ast} = 0.4$ and in units of $\mu_0 \simeq 1~ {\rm MeV}$.

%ADD TABLE?

Note that within a few orders of magnitude in energy, $\alpha$ itself has changed by more than an order of magnitude.  In the following we set the particle/antiparticle asymmetry to zero, but note that relaxing this assumption modifies thermal freeze-out in important ways~\cite{Graesser:2011wi} (see also~\cite{Griest:1986yu,Belyaev:2010kp}).  This therefore underscores the importance when comparing high- and low-energy DM processes. The scale $\mu_{S}$ allows, de facto, a clean separation between two distinct physical regimes for our DM theory.

In the following we will assume that either $\alpha_{X}$, $\alpha_{q}$ or both are asymptotically safe couplings. We shall refer to this scenario as asymptotically safe DM (asDM).

%%%%%%%%%
%\begin{figure}[b!]
%\includegraphics[width=.45\textwidth]{LambdaAF_v3}
% \caption{Dependence of the intrinsic energy scale $\Lambda_{AF}$ on $b_1$ for $\alpha^{\ast} = 0.4$ and $\alpha(\mu_0)=0.04$ for $\mu_0 = 1$ MeV. Above $\Lambda_{AF}$ the coupling has already reached over $2/3$ of its fixed point value. }
%\label{LambdaAF}
%\end{figure}
%%%%%%%%%% 
 
%%%%%%%%%%%%%%%%%%%%%%%%%%%%%%%
\begin{figure*}[t] %  figure placement: here, top, bottom, or page
\begin{center}
 \includegraphics[width=.32\textwidth]{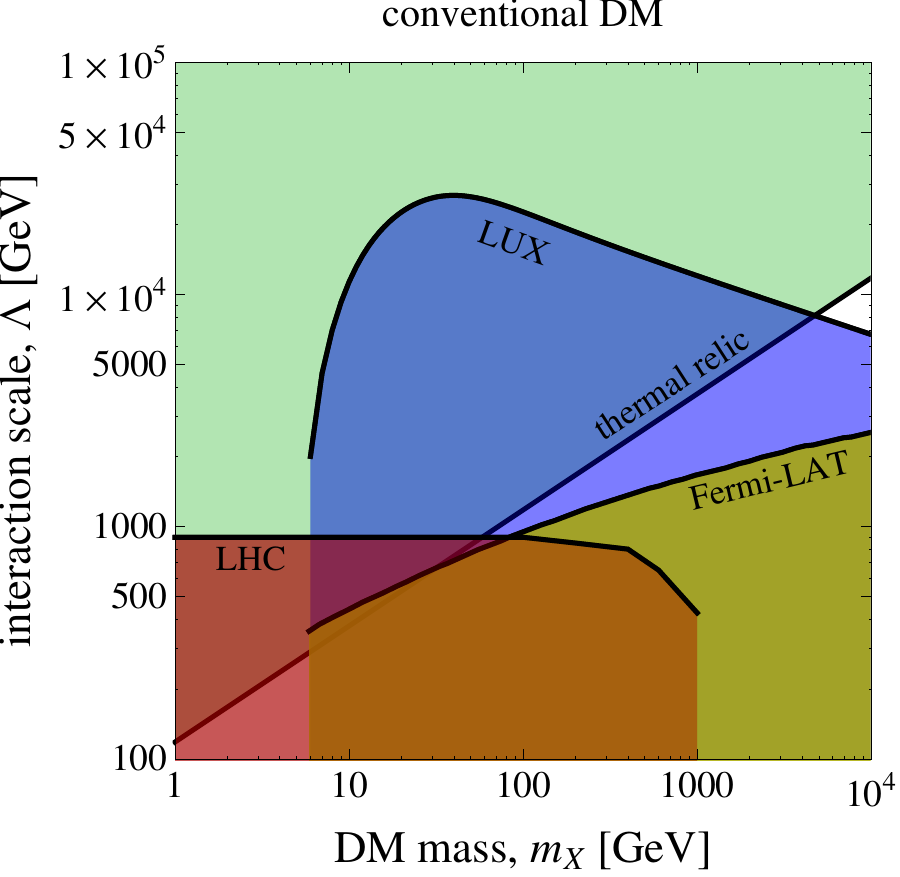} 
  \includegraphics[width=.32\textwidth]{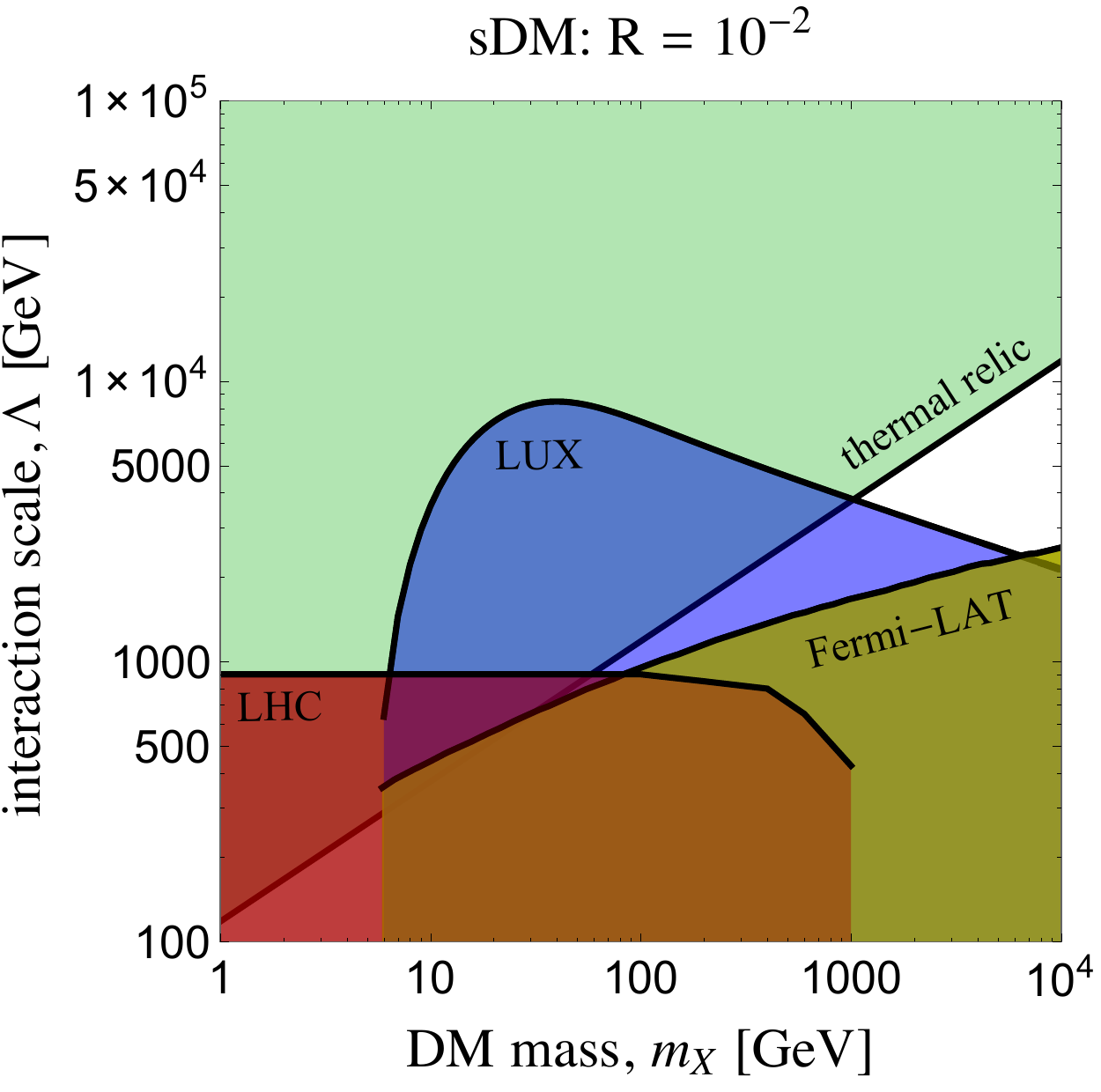}  
 \includegraphics[width=.32\textwidth]{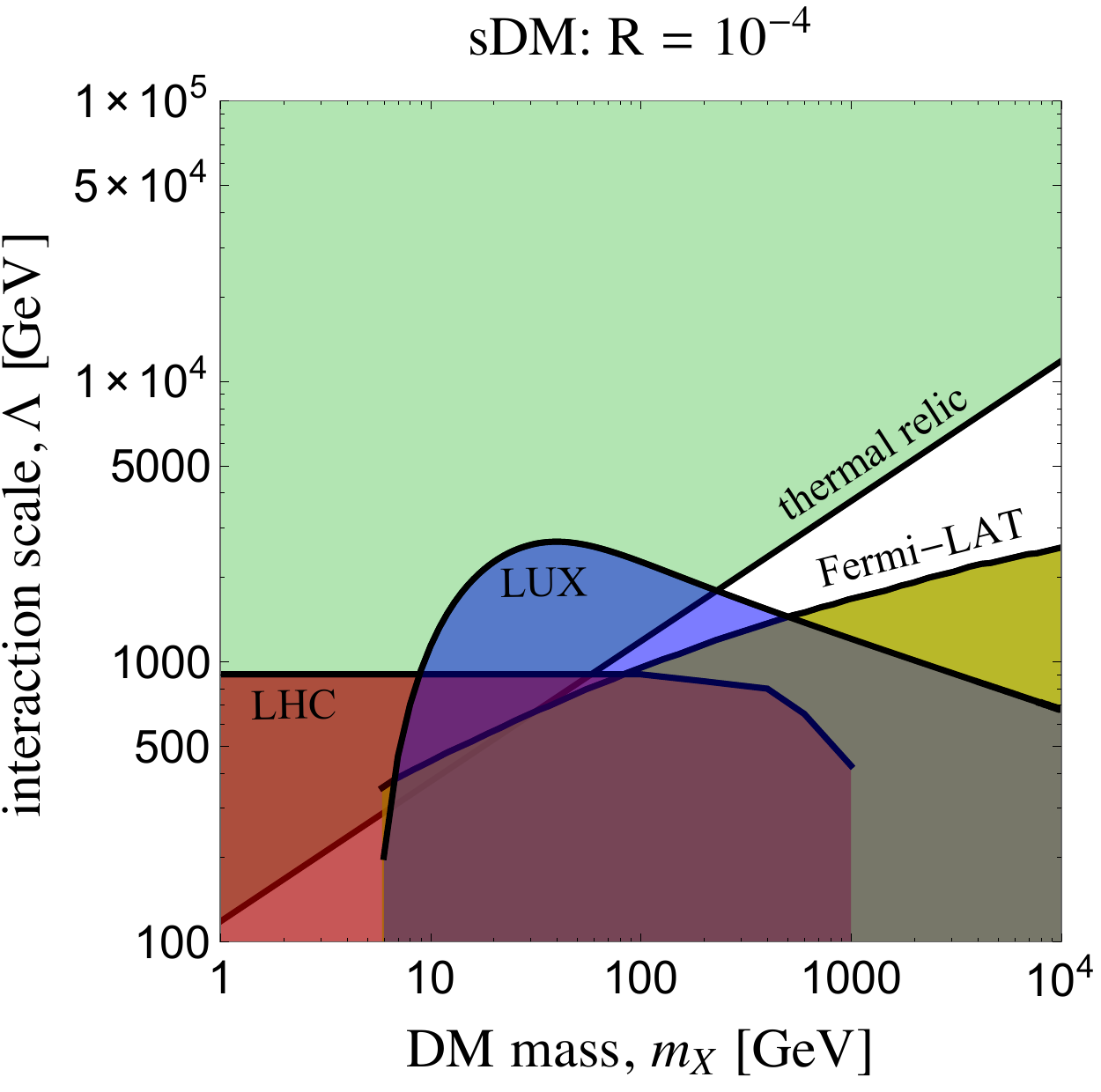}  
\caption{Here we illustrate the combined impact of various experimental probes on the relic abundance {{for the model described by Eq. (5).  The left panel illustrates the present status of ``conventional DM" interacting with quarks {\it without} sizeable running, as is typically assumed. The middle and right panels illustrate the impact of asymptotically safe couplings on easing the tension between direct detection and thermal relic requirements.}} The parameter, $R$ defined in Eq.~(\ref{eq:R}), parameterizes the difference between the low- and high-energy values of the couplings.  The experiments depicted include LHC monojets~\cite{CMSmonojets}, direct detection constraints from LUX~\cite{Akerib:2013tjd}, and Fermi-LAT's dwarf galaxy search~\cite{Ackermann:2013yva}.  The white space in each panel represents the remaining viable parameter space for a thermal relic. }
\label{fig:mav}
\end{center}
\end{figure*}

%%%%%%%%%%%%%%%%%%%%%%%%%%%%%%%
%------------------%
%\section{Phenomenology}

%------------------%

\textbf{\textit{Phenomenology -}} Let us now investigate first the consequences of asDM for the thermal relic abundance. In particular, we will focus on the phenomenologically interesting case in which the transition scale {$\mu_{S}$} is smaller than the freeze-out temperature but higher than the direct detection energy scale which is of $\mathscr{O}({\rm MeV})$.  In this case, freeze-out occurs when the coupling is nearly maximal and the direct detection experiments probe only very tiny asDM couplings. This typically means 
\be 
{\rm MeV} \lesssim \mu_{S} \lesssim \frac{m_{X}}{20},
\ee
and that direct detection constraints are weakened without negatively impacting the requirements of a thermal relic. We have confirmed numerically that this is an excellent approximation (see the right panel of Fig.~\ref{betafig}). 

We now briefly consider the case in which the mediator is instead much lighter than the DM in Eq.~\ref{eq1}.  An interesting application in this case, is this possibility that the first hints of asDM may already be seen in the Galactic Center excess (GCE)~\cite{Goodenough:2009gk,Abazajian:2012pn,Gordon:2013vta,Macias:2013vya,Abazajian:2014fta,Daylan:2014rsa,Calore:2014xka}.  Constructing a viable DM model for the GCE has proven challenging given the strong null observations in direct searches. Although we will consider a DM interpretation, we urge caution in interpreting the GCE and note that an astrophysical explanation may yet account for the signal~\cite{OLeary:2015gfa}.  Some viable DM explanations that have been explored include DM-SM interactions via pseduo-scalar exchange~\cite{Boehm:2014hva,Berlin:2014tja,Ipek:2014gua,Arina:2014yna,Dolan:2014ska} or ``flavored DM''~\cite{Agrawal:2014una,Izaguirre:2014vva} which suppress direct detection rates by employing spin- and momentum-dependent interactions or flavor-dependent couplings respectively.  Safe DM allows for a new qualitatively distinct class of models to account for the GCE in a viable way. We display one illustrative fit to the GCE in the right panel of Fig.~\ref{betafig}, where we have fixed the DM mass to 26.7 GeV~\cite{Calore:2014nla} which is the best-fit mass assuming equal coupling to all quark flavors.

It is important to stress, however, that both high-energy collider and indirect annihilation searches probe the large couplings of asDM at high energies. Let us illustrate this via a Maverick DM model~\cite{Beltran:2010ww}. In these models DM is a Maverick in the sense of being the only light particle associated with the dark sector feeling the SM fields. Thus, the particle mediating the interactions between DM and the SM are so heavy that their effects can be parameterized by an effective operator. Next, we will illustrate the impact of dark asymptotically safe couplings in the case of a heavy vector exchange between DM and quarks: 
\be 
\mathscr{O}_{V} = \frac{1}{\Lambda^{2}} \left(\overline{X} \gamma^{\mu}X\right) \left( \overline{q} \gamma_{\mu}q\right), 
\ee
where the scale of the operator can be matched onto a UV description via $\Lambda \equiv m_{V}/\sqrt{g_{X}g_{q}} \equiv \frac{m_{V}}{\sqrt{4\pi} (\alpha_{X}\alpha_{q})^{1/4}}$, where $m_{V}$ is the mass of the heavy vector and $g_{X}, g_{q}$ are the couplings to asDM and quarks respectively.

Next we determine the values of $\Lambda$ that satisfy the relic abundance by solving the Boltzmann equations, 
\be \frac{dn_{i}}{dt} + 3 H n_{i} = - \langle \sigma_{ann} v_{rel} \rangle \left[ n_{i}n_{j} - n_{eq}^2\right],
\label{eq:boltz}
\ee
where the indices run over $i,j = X, \overline{X}$. $H$ is the Hubble expansion rate, $n_{eq}$ is the equilibrium number density, and $\langle \sigma_{ann} v_{rel} \rangle$ is the thermal average of the total annihilation cross section. For the operator $\mathscr{O}_{V}$ the annihilation cross section is simply
\bea
\label{eq:relic}
& \langle \sigma_{ann} v_{rel} \rangle %&\\ 
%&=&
 =\frac{3m_{X}^{2}}{2\pi \Lambda_{*}^{4}} \sum_{q} \sqrt{1-\frac{m_{q}^{2}}{m_{X}^{2}}}\left(2+ \frac{m_{q}^{2}}{m_{X}^{2}}\right), %\nonumber
\eea
where we have neglected to include sub-leading $\mathscr{O}(v^{2})$ corrections, and the parameter  $\Lambda_{\ast}$ indicates the interaction scale when the couplings are near their fixed point value $\alpha^{\ast}_X \alpha^{\ast}_q$. {{Consistent with previous work, we find that for DM masses $\gtrsim 10$ GeV the requisite annihilation cross does not depend sensitively on its mass and is approximately $\langle \sigma_{ann}v_{rel}\rangle \simeq 6 \times 10^{-26}~{\rm cm^{3}}{\rm s}^{-1}$~\cite{Steigman:2012nb}}. When this is the case, the correct relic abundance requires $\Lambda_{\ast} \simeq 980~{\rm GeV} \left(\frac{m_{X}}{100~{\rm GeV}}\right)^{1/2}$, or equivalently at the level of couplings
\be \alpha_{X}^{*} \alpha_{q}^{*} = 7 \times 10^{-3} \left(\frac{m_{V}}{1~{\rm TeV}}\right)^{4} \left(\frac{100~{\rm GeV}}{m_{X}}\right)^{2} \ . 
\ee
To determine the direct detection cross section we should be able to run the couplings to lower energies. It could be that one or both couplings run to a lower value with decreasing energies. Introducing the effective direct detection interaction scale 
\begin{equation}
\Lambda_{DD} = \Lambda_{\ast} \left( \frac{\alpha_X^{\ast}\alpha_q^{\ast}}{\alpha_X \alpha_q}\right)^{\frac{1}{4}}  = \Lambda_{\ast}  \left(R_X R_q \right)^{-\frac{1}{4}}  \ , 
\label{eq:R}
\end{equation}
with $R_{X,q} = \alpha_{X,q}/\alpha_{X,q}^{\ast}$ and $ \alpha_{X,q}$ the couplings at the relevant  direct detection energies of order $\mathscr{O}({\rm MeV})$. We can now proceed with writing down the associated direct detection cross section
\be 
\label{eq:DD}
\sigma_{p} = \frac{\mu_{p}^{2}}{\pi \Lambda_{DD}^{4}} =\frac{\mu_{p}^{2}}{\pi \Lambda_{\ast}^{4}}R \ ,
\ee
where $\mu_{p}$ is the proton-DM reduced mass and $R \equiv R_X R_q$. For illustration we set $R = 10^{-2}$ and $10^{-4}$ and display the resulting constraints in the right panel of Fig.~\ref{fig:mav}.

Now we would like to roughly estimate the parameters of interest that suppress direct detection constraints enough to allow for viable thermal relics. Combining Eqs.(\ref{eq:relic}-\ref{eq:DD}) we see that we need 
\bea 
&&\frac{\Lambda_{DD}}{\Lambda_{\ast}}> %\\
%&&
 18 \left(\frac{100~{\rm GeV}}{m_{X}}\right)^{1/2} \left(\frac{1.3 \times 10^{-45}~{\rm cm^{2}}}{\sigma_{p}}\right)^{1/4} %\nonumber
\eea
where we have taken $\sigma_{p}$ to be below the constraints imposed by LUX~\cite{Akerib:2013tjd} which implies 
\begin{equation}
R < 10^{-5} \left(\frac{m_{X}}{100~{\rm GeV}}\right)^{2} \left(\frac{\sigma_{p}}{1.3 \times 10^{-45}~{\rm cm^{2}}}\right) \ .
\end{equation}

%We now briefly consider the case in which the mediator is instead much lighter than the DM.  An interesting application in this case, is this possibility that the first hints of sDM may already be seen in the Galactic Center excess (GCE)~\cite{Goodenough:2009gk,Abazajian:2012pn,Gordon:2013vta,Macias:2013vya,Abazajian:2014fta,Daylan:2014rsa,Calore:2014xka}.  Constructing a viable DM model for the GCE has proven challenging given the strong null observations in direct searches. Although we will consider a DM interpretation, we urge caution in interpreting the GCE and note that an astrophysical explanation may yet account for the signal~\cite{OLeary:2015gfa}.  Some viable DM explanations that have been explored include DM-SM interactions via pseduo-scalar exchange~\cite{Boehm:2014hva,Berlin:2014tja,Ipek:2014gua,Arina:2014yna,Dolan:2014ska} or ``flavored DM''~\cite{Agrawal:2014una,Izaguirre:2014vva} which suppress direct detection rates by employing spin- and momentum-dependent interactions or flavor-dependent couplings respectively. Asymptotically safe DM allows for a new qualitatively distinct class of models to account for the GCE in a viable way.  

%The dwarf limits of Ref.~\cite{Ackermann:2013yva} combines Fermi-LAT observations in the directions of a total of $15$ satellites galaxies of the Milky Way, and are an update of previous estimates \cite{Abdo:2010ex,GeringerSameth:2011iw,Ackermann:2011wa} in both data and analysis

%\section{Model building and Conclusions}
\textbf{\textit{Model Building and Conclusions -}} To construct an asDM model we take inspiration from and make use of, the exact results in \cite{Litim:2014uca}. Here it was proven the existence of a gauge-Yukawa theory, structurally similar to the SM, featuring a one-dimensional critical RG hypersurface in the four-dimensional coupling space along which the physical theory runs from a sensible IR non-interacting field theory to a quantum interacting UV fixed point.
The fact that the hypercritical surface is unidimensional means that along the RG trajectory connecting the IR and UV physics all the couplings display asymptotically safe behavior and all the couplings are driven by only one coupling, which is in this case the gauge coupling. We will assume that an underlying theory similar to this this is the dark sector driving the running of sDM couplings to itself and to ordinary matter. A similar construction was considered in \cite{Abel:2013mya}, albeit in a different context. The hidden theory is constituted by an $SU(N_h)$ gauge theory featuring $F_{ h}$ hidden Dirac fermions $\psi_{h}$ in the fundamental representation and interacting among themselves via a complex matrix of $F_{ h} \times F_{h}$ scalars. The ratio of the number of hidden flavors to hidden colors is chosen in such a way that asymptotic freedom is lost. The same ratio is also the parameter used to control and insure the presence of an exact interacting UV fixed point within perturbation theory \cite{Litim:2014uca}. We will indicate the Lagrangian of this sector collectively with $\mathscr{L}_{\rm hidden}$. We assume that our sDM state is one extra heavy Dirac flavor $X$, with an exact unbroken flavor symmetry. We furthermore assume that at energies higher than the mass of $X$ the full hidden symmetry gauge group is $SU(N_h + 1 )$. Similarly the nonabelian hidden global symmetry is $SU(F_h +1) \times SU(F_h +1)$. Both, the hidden gauge and global symmetries, spontaneously break at around the vector mass scale $m_V$  while we keep $m_X < m_V$. At energies below and near $m_V$ the physics is well-described by Eq.~(\ref{eq1}).
%\begin{eqnarray}
%\mathscr{L} &=& \mathscr{L}_{\rm hidden} + i\bar{X} \gamma^{\mu}\left(\partial_{\mu} - i g_X V_{\mu}  \right) X \nonumber \\ &&+ \bar{q} \gamma^{\mu}\left(\partial_{\mu} - i g_q V_{\mu}  \right) q + m^2_V V_{\mu}V^{\mu}\ . 
%\end{eqnarray}
Here $V_{\mu}$ is an abelian massive vector field that is part of the larger gauge symmetry group and we neglected its kinetic term. We further assume it to couple universally also to the SM quarks. At some higher energies we can imagine an unification also with the SM fields, provided that it still leads to an asymptotically (near) safe behavior for either or both sDM relevant couplings $g_X$ and $g_q$. With this setup   at energy scales below $m_V$  the hidden sector drives the running of, at least, $g_X$. By the fundings in \cite{Litim:2014uca} the cartoon beta function responsible for the running in Fig.~\ref{betafig} maps into in the beta function in Fig.~5 of \cite{Litim:2014uca}. 

The running of $g_X$ above the $X$ and $V_{\mu}$ mass thresholds should be amended by enlarging the hidden color and flavor group, which by construction is structurally identical to the theory with one less hidden color and flavor and therefore we expect the UV ultraviolet fixed point to survive, at least within the energy range relevant for sDM phenomenology. 

Although the results in \cite{Litim:2014uca} are exact in the Veneziano limit, for phenomenological reasons, we extend them to finite number of hidden flavors and colors. Here the beta function for $\alpha_X$, after having already zeroed the Yukawa beta function, maps into Eq.~\eqref{ASFBeta} for
\begin{equation}
 \quad b_1 = \left(\frac{N_c}{4\pi}\right)^2 \left(\frac{50}{3} - \frac{8}{3} \frac{N_f}{N_c} + \frac{6 N_c}{N_c + N_f }\right)
 \end{equation}
and the fixed point value of the gauge coupling : 
\begin{equation}
\alpha_X^{\ast} \simeq \frac{N_c}{4\pi} \frac{4}{ 3 b_1 }\left(\frac{N_f}{N_c} - \frac{11}{2} \right) \ .
\end{equation}
With $N_c = N_h+1$ and $N_f = F_h +1$, with $N_c$ and $N_f$ large and $N_f/N_c - 11/2 < 1$.  
Choosing, for example, $N_h = 39$ and $F_h = 34$ we have $\alpha_X^{\ast} \sim 0.76$. One finds that $\Lambda_{\ast} \simeq 1.75$~GeV for $\alpha_{X}(1~{\rm MeV}) \simeq 0.04$. Larger values of $\alpha_X^{\ast}$ are obtained by decreasing  $N_f/N_c$ towards $11/2$.   
Note that in this extreme case $\alpha_X^{\ast}/\alpha_{X}(1~{\rm MeV}) \simeq 19$.

%%%%%%%%%%
%\section{conclusions}
%%%%%%%%%%%
We have shown that the interaction strength of DM interactions need not be constant with energy, and investigated the consequences of asymptotically safe couplings for the thermal relic abundance. We have observed that the running of the couplings can be very relevant when the transition energy scale falls in between the low-energy scale relevant for direct detection and the relatively high scales relevant for thermal freeze-out. By suppressing the otherwise extremely strong constraints from direct detection, the constraints from collider and indirect searches increase in importance. Although we have focused on the consequences of asymptotically safe couplings for symmetric thermal relics, it would be natural to extend this analysis to asymmetric thermal relics by making use of the indirect limits obtained in~\cite{Bell:2014xta}.
\vskip .2cm 
The CP$^3$-Origins center is partially funded by the Danish National Research Foundation, grant number DNRF90.

{\section*{Note added} A related study recently appeared~\cite{Davoudiasl:2015hxa}. They study the use of fixed target experiments for probing the running of dark couplings. The model employed therein can be viewed as a UV ``un-safe'' limit of the approach adopted in the present paper.}

\bibliography{nu}

\end{document}